\documentclass[prd,twocolumn]{revtex4}
\usepackage{graphicx}
\usepackage{amssymb}
\usepackage{amsmath}
\usepackage{epstopdf}

\newcommand{\lya}{Lyman-$\alpha$ }

\begin{document}

\title{The Acoustic Peak in the Lyman Alpha Forest}
\author{An\v{z}e Slosar}
\email{anze@berkeley.edu}
\affiliation{Berkeley Center for Cosmological
  Physics, Physics Department and Lawrence Berkeley National
  Laboratory,University of California, Berkeley California 94720, USA}
\author{Shirley Ho}
\email{cwho@lbl.gov}
\affiliation{Lawrence Berkeley National Laboratory, Berkeley, CA 94704}
\author{Martin White}
\affiliation{Department of Physics and Astronomy, University of California Berkeley, CA 94720}
\author{Thibaut Louis}
\affiliation{Ecole Normale Sup\'{e}rieure, D\'{e}partement de Physique, 24 rue Lhomond
75005 Paris, France
}

\date{\today}

\begin{abstract}
  We present the first simulation of the signature of baryonic
  acoustic oscillations (BAO) in \lya forest data containing 180,000
  mock quasar sight-lines.  We use eight large dark-matter only
  simulations onto which we paint the \lya field using the fluctuating
  Gunn-Peterson approximation. We argue that this approach should be
  sufficient for the mean signature on the scales of interest. Our
  results indicate that \lya flux provides a good tracer of the
  underlying dark matter field on large scales and that redshift space
  distortions are well described by a simple linear theory
  prescription.  We compare Fourier and configuration space approaches
  to describing the signal and argue that configuration space
  statistics provide useful data compression. We also investigate the
  effect of a fluctuating photo-ionizing background using a simplified
  model and find that such fluctuations do add smooth power on large
  scales. The acoustic peak position is, however, unaffected for small
  amplitude fluctuations ($<10\%$).  Larger amplitude fluctuations
  make the recovery of the BAO signal more difficult and  may degrade 
  the achievable significance of the measurement.
\end{abstract}

\maketitle

\section{Introduction}

Oscillations of the baryon-photon plasma in the early universe, also
known as Baryon Acoustic Oscillations (BAO), imprint a distinct
signature on the clustering of matter
\citep{1970ApJ...162..815P,1970Ap&SS...7....3S}.  The distance that
acoustic waves propagate in the first several hundred thousand years
of cosmic evolution set a characteristic scale that is measurable as a
distinct peak in the correlation function of matter fluctuations, or
as an oscillatory pattern in the power spectrum of the same (see
\cite{1998ApJ...496..605E,1999MNRAS.304..851M} for a detailed
description of the physics in modern cosmologies and
\cite{2007ApJ...664..660E} for a comparison of Fourier and
configuration space pictures).  These oscillations have been
traditionally measured in the Cosmic Microwave Background (see
\cite{2009ApJS..180..296N} for the latest) but with advent of new,
large-volume galaxy redshift surveys BAO have been detected in galaxy
clustering at low $z$ as well
\cite{eistenstein05,2005MNRAS.362..505C,2006A&A...449..891H,2007MNRAS.374.1527B,padmanabhan07,2007MNRAS.381.1053P}.

The use of BAO as a probe of cosmological parameters is especially enticing
since the signal is at relatively large scale (around $150\,$Mpc), where the
modes are still mostly in the linear regime.  The power spectrum or correlation
function can be thus computed quite accurately with only linear perturbation
theory once one specifies the baryon-to-photon ratio and matter-radiation ratio,
which are both measured accurately from CMB acoustic peaks
\cite{1997PhRvD..56..596H,2002AmJPh..70..106W}.
The ability to calibrate the BAO signal provides a standard ruler in both
the transverse and radial directions, allowing one to measure the the angular
diameter distance and the Hubble parameter as a function of redshift in the
clustering of matter.
Absent any systematic errors, obtaining a high precision measurement of the
distance simply requires surveying a large volume and locating the features
in the 2-point function corresponding to the acoustic scale.
The recent measurements of the distance scale to $z\simeq 0.2-0.4$ provide
us with complementary constraints to other large scale structure probes,
significantly improving constraints on key cosmological parameters.

The BAO technique becomes even more powerful as one moves to higher redshift,
where the acoustic scale is expected to be more linear and at which more
volume is available to be surveyed.  Unfortunately tracing the large volumes
with high fidelity, as is required by BAO studies, becomes increasingly
expensive of telescope time if galaxies are used as the tracer.
However, in principle any tracer of the mass field will do, including the
neutral hydrogen in the inter-galactic medium (IGM) \cite{2002MNRAS.329..848V,2003dmci.confE..18W}
or galaxies \cite{2008PhRvL.100i1303C}.  Tracing neutral hydrogen in galaxies
via its redshifted $21\,$cm emission is a key goal for proposed future radio
telescopes.  However, even with current technology it is relatively
straightforward to obtain a low resolution spectrum of distant quasars and
study the \lya forest of absorption lines which map the neutral hydrogen along
the line-of-sight.
At $z\simeq 2-3$ the gas making up the IGM is in photo-ionization equilibrium,
which results in a tight density-temperature relation for the absorbing
material with the neutral hydrogen density proportional to a power of the
baryon density \cite{1997MNRAS.292...27H,2007arXiv0711.3358M}.
Since pressure forces are sub-dominant, the neutral hydrogen density closely
traces the total matter density on large scales.
The structure in QSO absorption thus traces, in a calculable way, slight
fluctuations in the matter density of the universe back along the line-of-sight
to the QSO, with most of the \lya forest arising from over-densities of a
few times the mean density.

The measurability of the BAO signal in the \lya forest has received relatively
little attention in the literature.
Motivated by this and the development of the upcoming BOSS experiment, which
will deliver unprecedented number of quasar spectra probing the \lya forest
at $z\sim 2-3$, we further develop the theory.
It is difficult to compute the BAO signal in the forest analytically,
involving as it does a projection of a non-linear mapping of a non-linear
density field in redshift space.  For this reason we resort to large N-body
simulations.
Our work is an expansion of, and is complementary to,
\cite{2007PhRvD..76f3009M} who estimated the potential of a survey such as
BOSS to measure the BAO scale assuming that the \lya flux traces the dark
matter on large scales, following prescriptions from \cite{2003ApJ...585...34M}.
This work confirms that intuition by an explicit modeling of the \lya forest
flux, albeit in a simplified manner that does not fully capture the
all of the small-scale physics.

The outline of the paper is as follows.  We describe our simulations
in Sec.~\ref{sec:sim}, discussing our results for the 2-point
statistics of the Lyman-$\alpha$ forest in Sec.~\ref{sec:signal}.  We
measure and attempt to explain the modification of the signal by
non-linear effects such as redshift-space distortions in the flux
decrement correlation function.  Furthermore, we make a preliminary
investigation of one of the many systematic effects that could change
the BAO signal -- UV background fluctuations -- in Sec.~\ref{sec:UV}.
Further investigation is underway to thoroughly understand the
systematic effects of the \lya forest BAO signal, but they are beyond
the scope of this paper and will be reported in a future publication.
Finally we discuss the possible strategies for the detection of the
BAO signal in the flux decrement correlation or powerspectrum and
conclude in Sec.~\ref{sec:conclude}.

\section{Simulations}
\label{sec:sim}

As mentioned above, it is difficult to compute the BAO signal in the \lya
forest analytically.  It is also a challenging problem numerically.
Since much of the signal of interest comes from near mean density gas,
mass resolution is as important as force resolution.
To resolve the Jean's scale of the gas $\mathcal{O}(100\,{\rm kpc})$ while
simultaneously simulating a representative volume stretches computational
abilities, even for gravity-only simulations.
We are thus forced to compromise.
As we argue below, the physics governing the small-scale fluctuations should
be approximately decoupled from the BAO scale, and inaccurate modeling
of the small-scale physics should lead to smooth modifications of the
flux power spectra or correlation functions which do not contain
imprints of the acoustic scale.

Guided by this reasoning we ran 8 particle-mesh simulations of a flat
$\Lambda$CDM cosmology, with $\Omega_M$ $=0.25$, $\Omega_\Lambda$
$=0.75$, $h=0.75$, $n=0.97$ and $\sigma_8$ $=0.8$.  Each simulation
evolved $3000^3$ particles in a $1500\,h^{-1}$Mpc box, computing the
forces on a $3000^3$ grid.  The particle data were dumped at $z=2.5$
and density and velocity fields generated from the particle positions
and velocities along a regular grid of $150^2=22,500$ lines-of-sight
using a spline kernel interpolation with an effective smoothing of
$250\,h^{-1}$kpc.  Though this smoothing is about twice the Jean's
scale at $z=2.5$, it is nearly three orders of magnitude smaller than
the BAO scale providing good scale separation between the mis-modeled
physics and the signal of interest.

For each line-of-sight the fluctuating Gunn-Peterson approximation
(FGPA,
\cite{1998ApJ...495...44C,1998MNRAS.296...44G,2007arXiv0711.3358M})
was used to generate skewers of optical depth with $3,000$ pixels
each.  We assumed a temperature at mean density of $2\times 10^4$ K
and an equation of state $\gamma$=$1.5$ \cite{2002ApJ...567L.103T}.
Different choices for the slope, even an inverted equation of state,
will quantitatively but not qualitatively change our conclusions.  The
optical depth included thermal broadening (assumed Gaussian) and
skewers were generated both with and without peculiar velocities for
the gas.  The optical depth was scaled so that the mean transmitted
flux $\bar{F}=\left\langle\exp(-\tau)\right\rangle=0.8$
\cite{2004MNRAS.350.1107M}.  For completeness, we also generate the
skewers with dark-matter over-density only, so we can compare the flux
statistics to those of the underlying mass.  We work throughout with
relative fluctuations in the flux, $\delta_F = F(\hat{x})/\bar{F} -1$,
so our fundamental data set is $\delta_F(\vec{x})$ on $150^2$ skewers
of $3,000$ pixels each. In the results presented in this paper,
$\bar{F}$ was determined globally for each box under consideration. We
have also attempted to determine $\bar{F}$ individually for each
skewer, which is closer to the observational situation where one has
fit continuum for each individual quasar sight-line. As expected, the
resulting two-point correlation functions changed by a small constant
offset, which can be easily modeled in the analysis of the real data.

In this exploratory work we neglect several higher order effects in order
to concentrate on the underlying physics.
First, we neglect the evolution of \lya forest with redshift -- both the
structure of the IGM and the mean flux -- and generate the skewers at $z=2.5$.
Second, we make the distant observer approximation that all skewers are
parallel when in fact the comoving radial distance changes by about 20\%
across the depth of our simulation.
This has two implications.
The first is that the transverse distance between two points is to within
$\sim 10\%$ of that obtained by assuming perfectly parallel skewers.
The second is that in reality the line-of-sight velocities are not perfectly
parallel and that the Kaiser formalism that we later employ will not be exact.
We will return to these issues in a future publication.

\begin{figure}
  \includegraphics[width=0.9\linewidth]{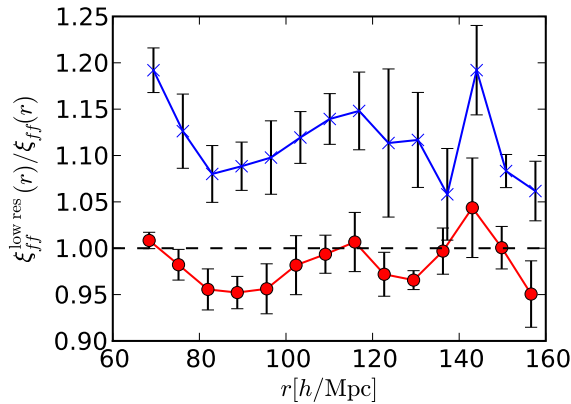} 
  \caption{The ratio of the monopole of the flux correlation function
    in redshift space for density fields smoothed on scales a factor
    of two (filled red points) or four (blue crosses) larger than our
    fiducial case.  Errors are heavily correlated and only meant to be
    indicative of the underlying uncertainties.}
    \label{fig:xires}x
\end{figure}

\begin{figure}
  \includegraphics[width=0.9\linewidth]{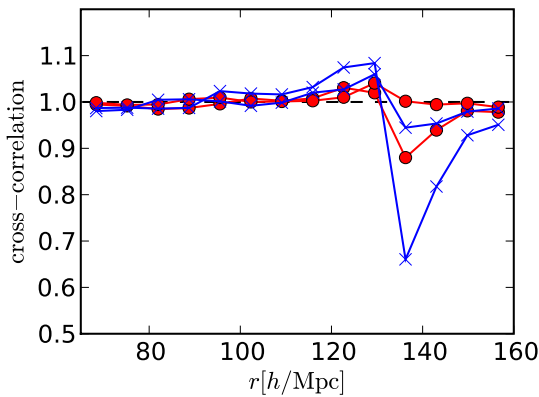} 
  \caption{The cross-correlation coefficient between the flux in our
    low and high resolution boxes,
    $\sqrt{\xi^2_{lh}/\xi_{ll}\xi_{hh}}$.  Filled red points show the result
    for the two low resolution boxes having twice the smoothing length
    of the high resolution box, blue crosses is the same for $4\times$
    smoothing length.}
  \label{fig:xirescc}
\end{figure}

There are two important effects that make our simulations inaccurate.
First, the simulations do not model the gas, assuming it faithfully traces
the dark matter on large scales (e.g.~\cite{2001MNRAS.324..141M}).
Even if we did attempt to model the gas, our resolution is not adequate to
model the small scale physics; a fully converged simulation would require a
resolution of $\sim 30\,h^{-1}\,$kpc, a factor of several smaller than we
achieve \cite{2007arXiv0711.3358M}.
Within the assumptions of the peak-background split the non-idealities will
be absorbed into smooth additive and multiplicative functions and will not
qualitatively alter our conclusions.

To test the effect of limited resolution on the BAO scale we smoothed the
3D density and velocity fields of two of our simulations by a factor of two
or four before computing the skewers.
As shown in Fig.~\ref{fig:xires} the simulations are quite well converged,
with the ratio of flux correlation functions agreeing to within 10\% for a
factor-of-two change in resolution.
We expect that the missing small scale physics will have a larger effect
on the covariance of the clustering, i.e.~the error bars, but we do not
currently have enough simulations to make this measurement reliably.

We can also compute the cross-correlation between the flux along the
same lines-of-sight in the two simulations with different 3D
smoothings of the density and velocity fields.  This directly tests
the impact of small-scale physics on how well the flux traces the
density on large scales.  Figure \ref{fig:xirescc} shows this cross
correlation.  We do not plot the errors, but the uncertainty can be
gauged from box-to-box scatter. The correlation is nearly unity on
large scales, indicating that the details of the small-scale physics
do not alter the behavior of the flux on large scales.  The signal
becomes difficult to measure beyond $100\,h^{-1}$Mpc where the
correlation functions are small and become negative.

\begin{figure*}
  \includegraphics[width=0.9\linewidth]{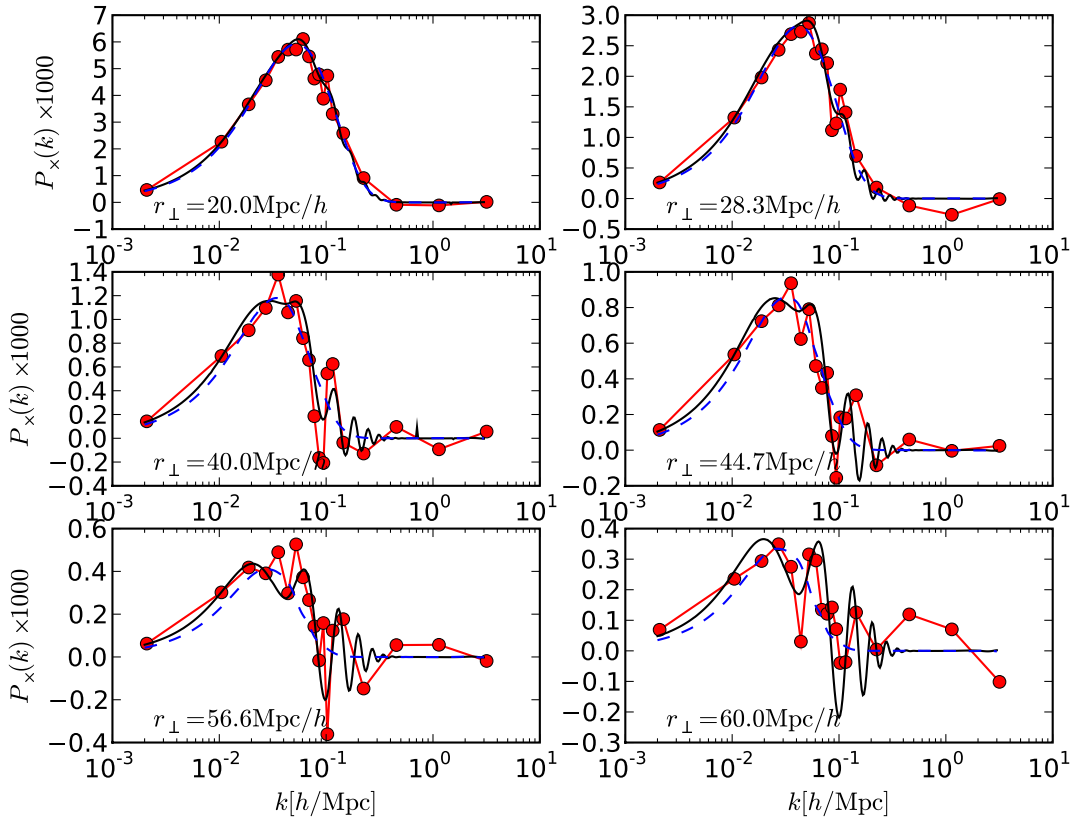} 
  \caption{The cross power spectrum for the matter overdensity.  Filled red
    points are measured from our simulations, the solid black line is the
    linear theory prediction, while the blue dashed line is the same for a
    baryon-less universe.}
    \label{fig:ps1}
\end{figure*}

\section{BAO signal}

\begin{figure*}
  \begin{tabular}{ccc}
    \includegraphics[width=0.3\linewidth]{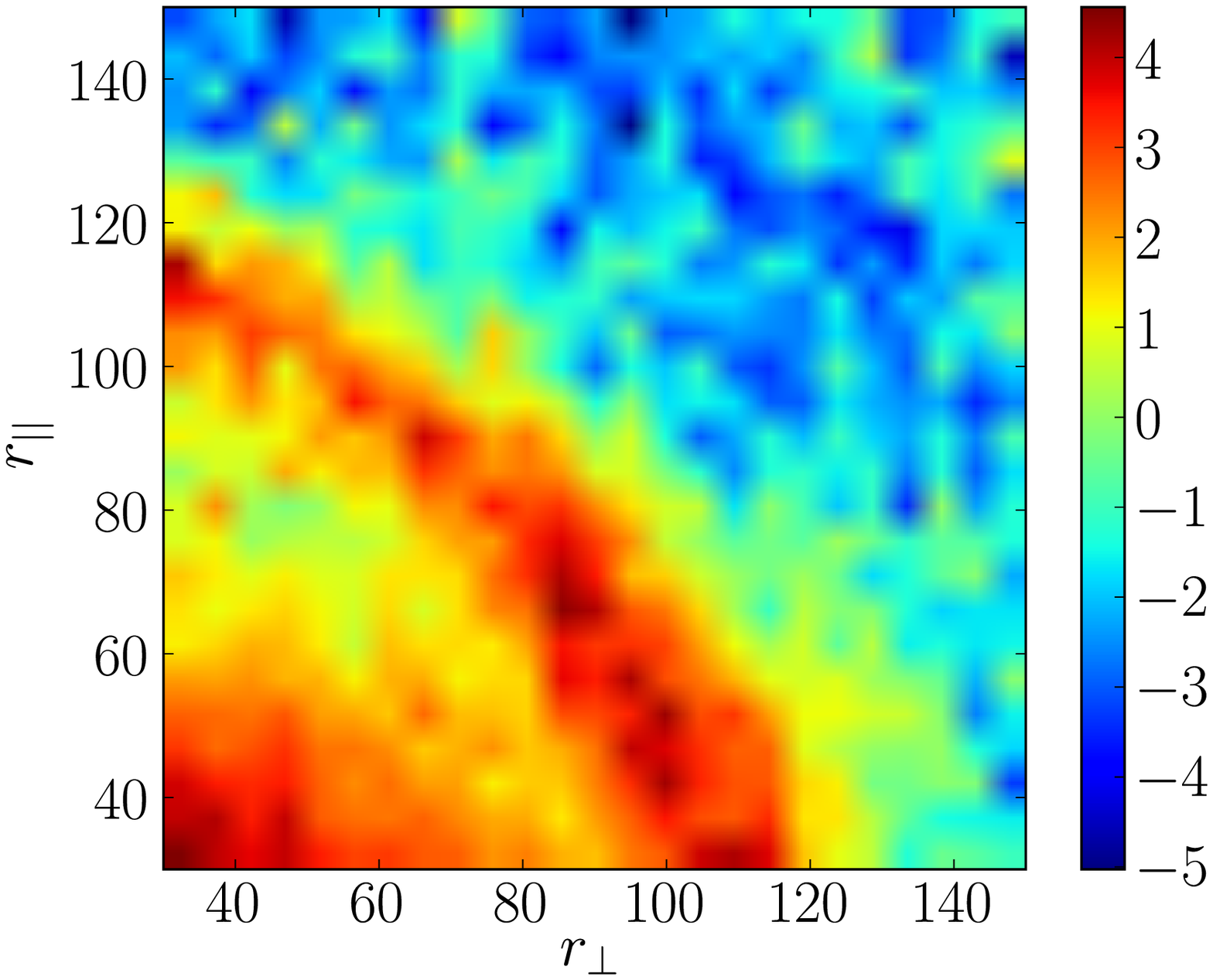} &
    \includegraphics[width=0.3\linewidth]{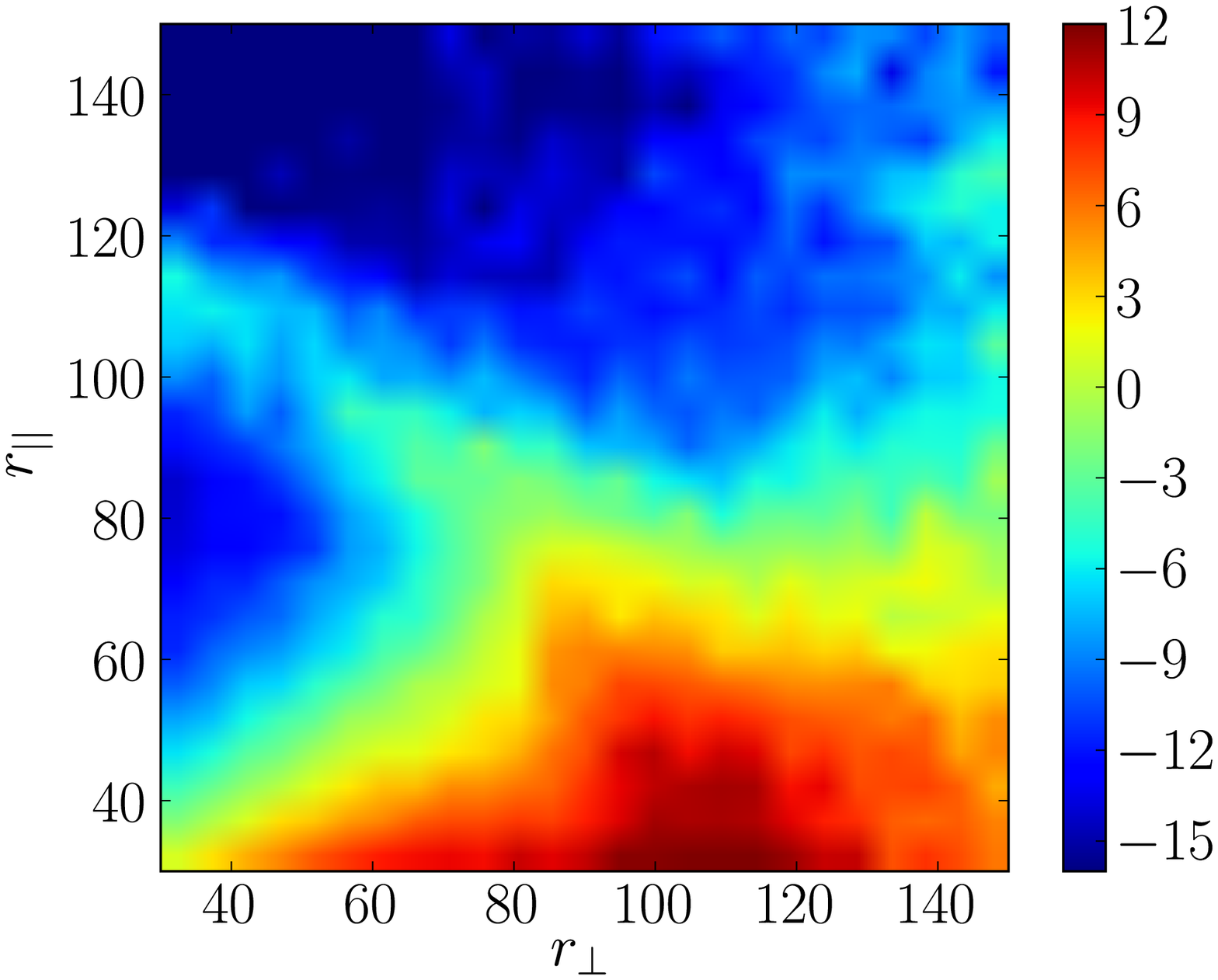} &
    \includegraphics[width=0.3\linewidth]{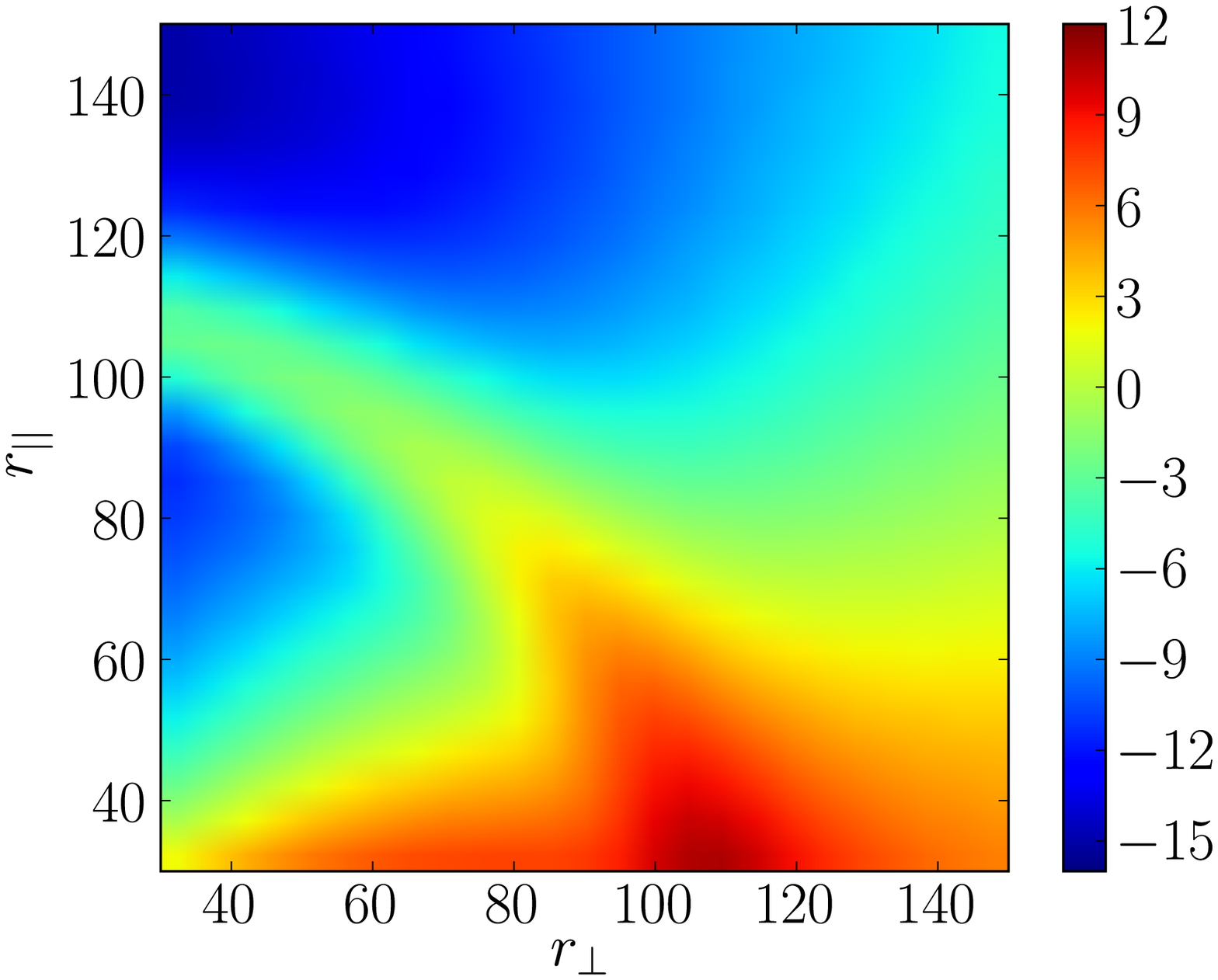} \\
    \includegraphics[width=0.3\linewidth]{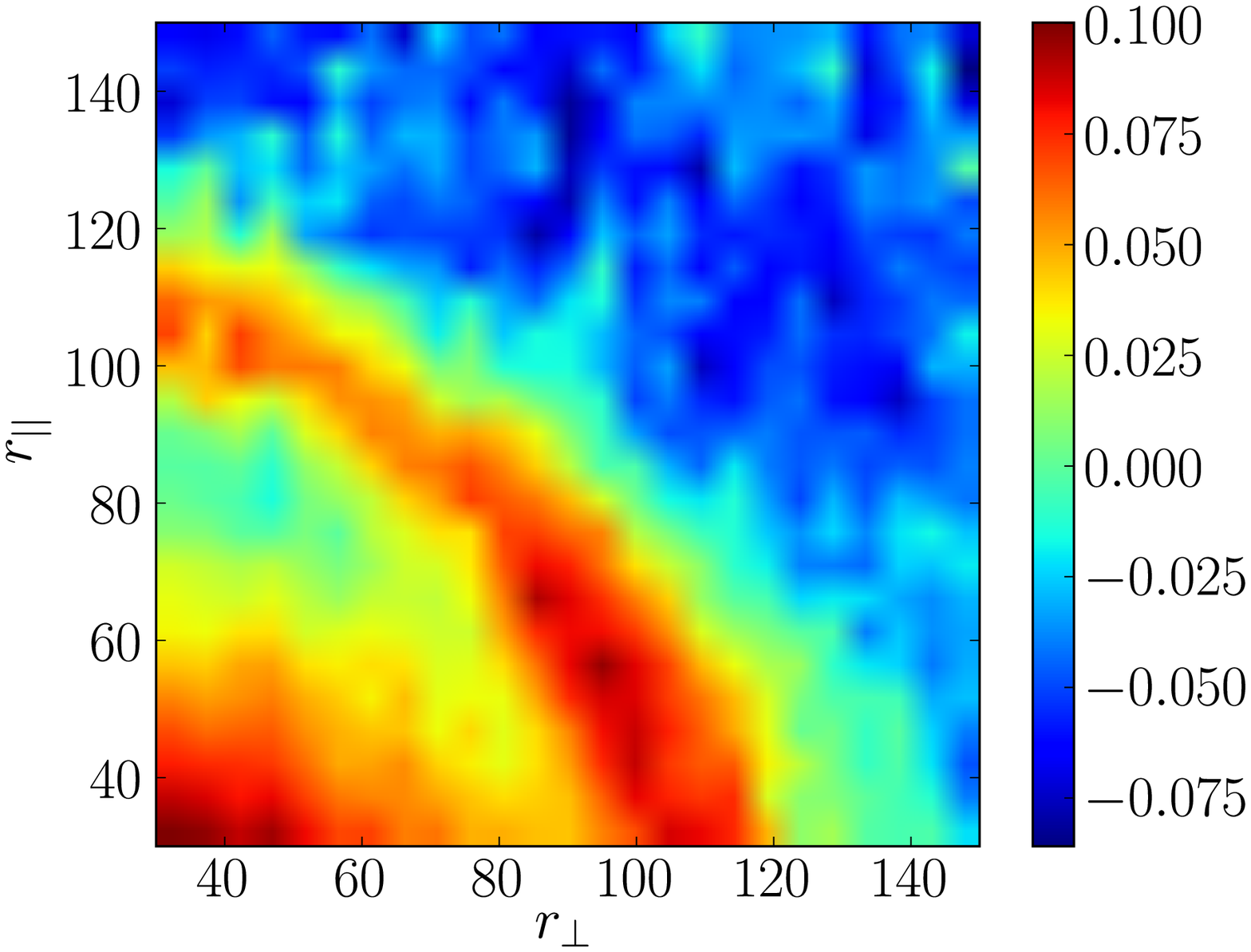} &
    \includegraphics[width=0.3\linewidth]{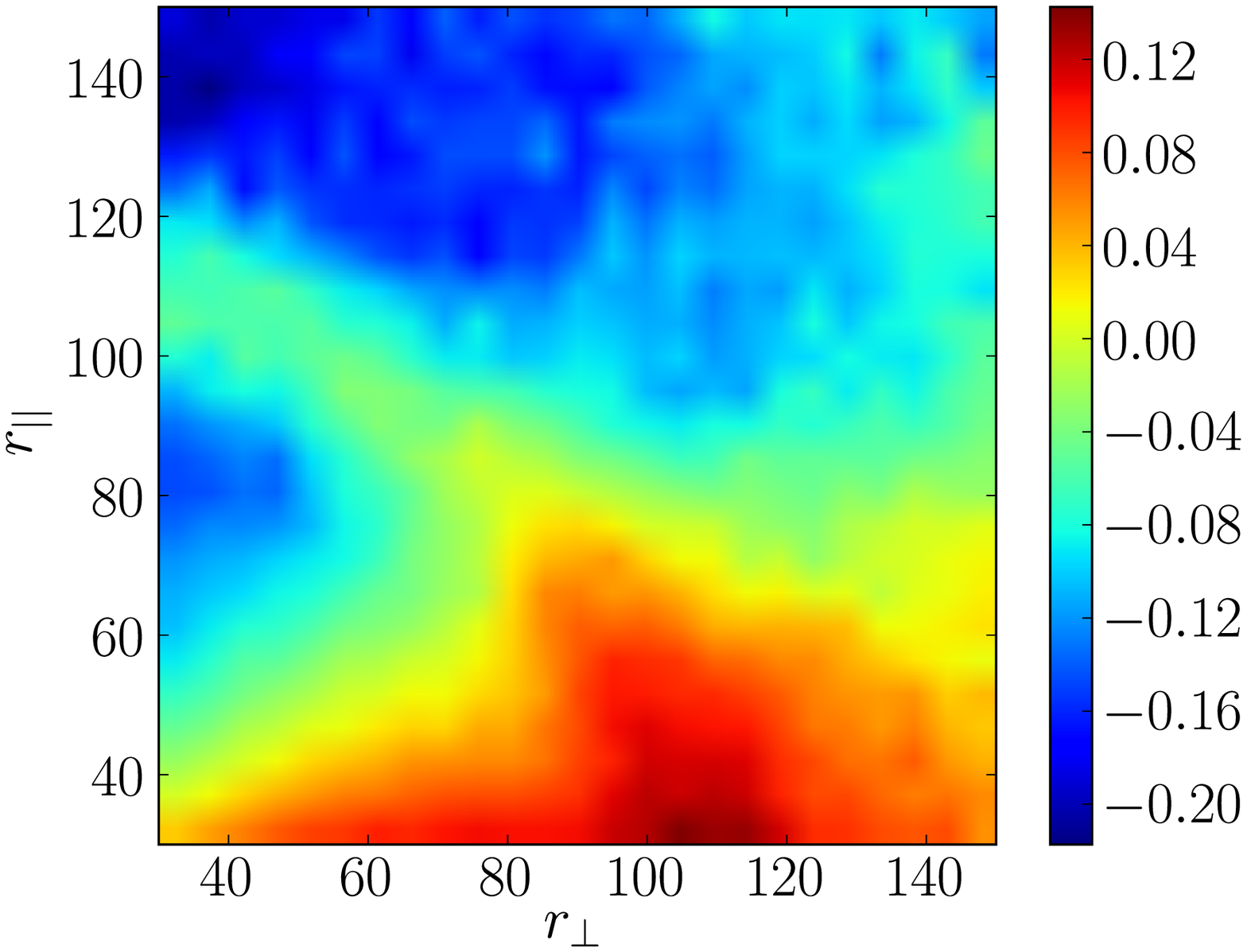} &
    \includegraphics[width=0.3\linewidth]{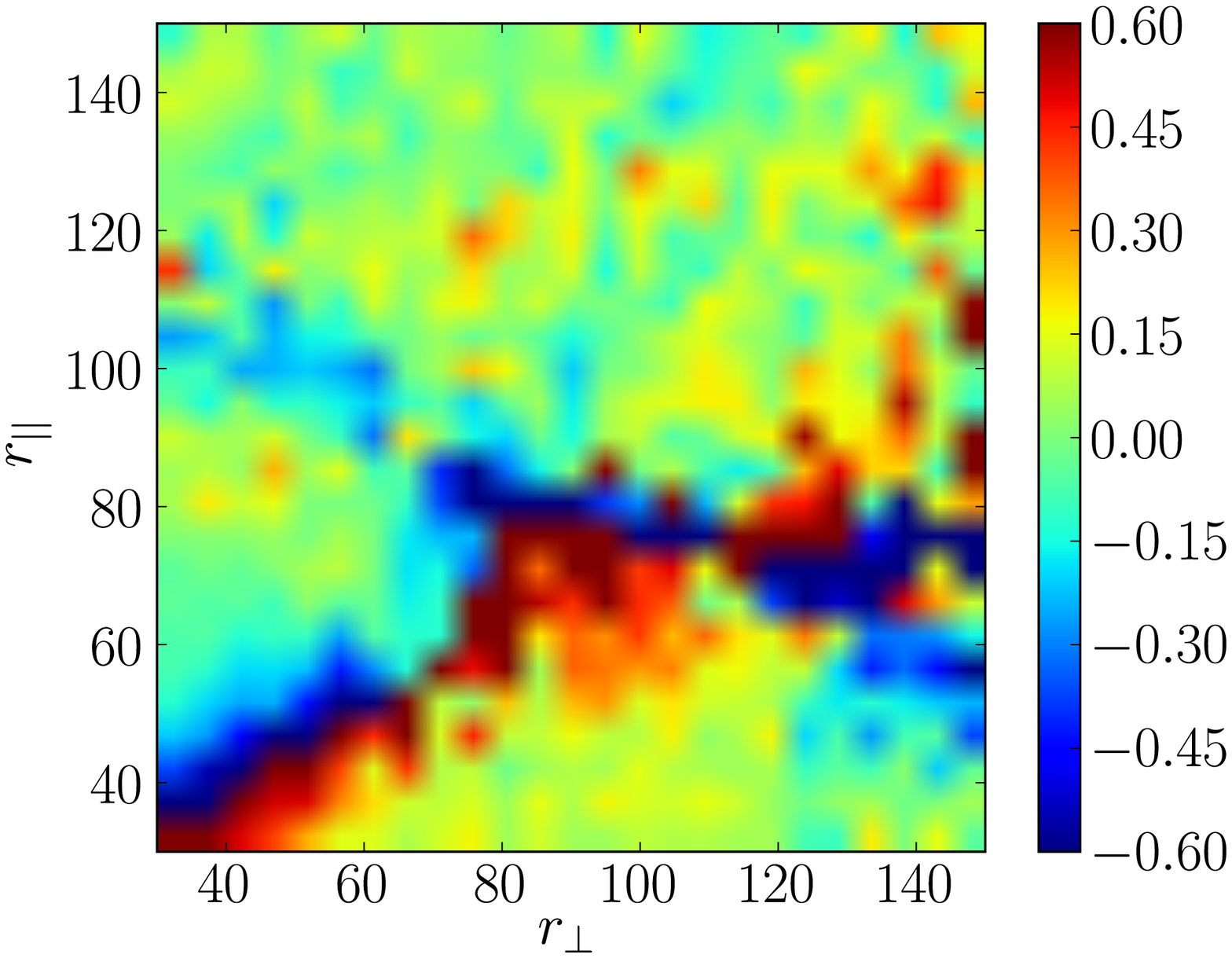} \\
  \end{tabular}
  \caption{The correlation functions, multiplied by $r^2$ for easier
    visualization, as a function of $r_\perp$ and $r_{||}$ in
    $h^{-1}$Mpc.  (Top) Correlation functions for matter over-density
    in real space, redshift space and the theoretical predictions for
    linear theory.  (Bottom) The correlation function in real and
    redshift space for the flux fluctuations, $\delta_F=F/\bar{F}-1$,
    and (bottom right) the relative difference between the flux and
    (appropriately scaled) dark matter correlation functions in
    redshift space $(\xi_{ff}-b^2\xi_{\delta\delta})/\xi_{ff}$. The
    features in this plot follow the crossing of $\xi_{ff}$ through
    zero, where the quantity we plot diverges.  Note that each panel
    has a very different color scale.}
    \label{fig:rxir}
\end{figure*}

\subsection{Fourier space}

We start by discussing the signal in Fourier space, where we expect
different $k$ modes to be decoupled in the linear regime.  Since the
signature of BAO in Fourier space is a set of oscillations rather than
a single peak, the aliasing of the modes introduced by a finite window
function becomes severe in our problem.  While this aliasing is most
well known from pencil-beam galaxy surveys, it becomes more even more
acute in the \lya forest as the data are even more sparsely sampled.
In fact, we were unable to obtain a measurement of the
three-dimensional flux power spectrum from our skewers, due to strong
mixing with the window function.  The regular distribution of our
skewers likely aggravated the problem, but it would remain even for
irregularly distributed skewers.

\begin{figure}
  \includegraphics[width=0.9\linewidth]{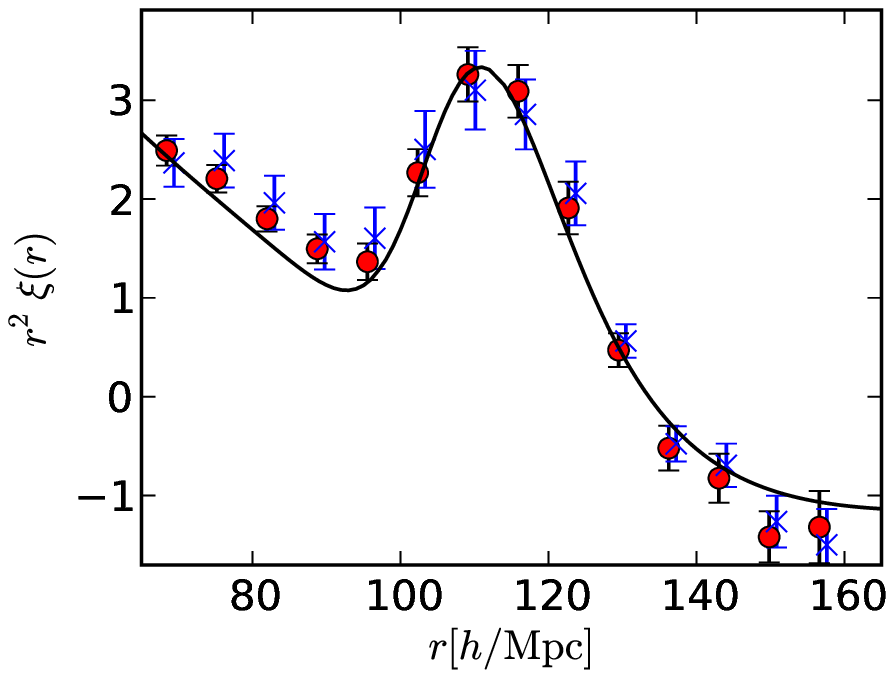} 
  \caption{The monopole correlation function of the dark matter, multiplied
    by $r^2$, in real space (filled red points) and redshift space (blue crosses),
    the latter divided by $C_0$.
    Error bars are estimated from the box-to-box scatter, are heavily
    correlated and only indicative of the underlying uncertainties.
    The linear theoretical prediction is plotted as the (black) solid line.
    \label{fig:xidd} }
\end{figure}

One can avoid the aliasing issue, however, if one works with cross-power
spectra, defined as
\begin{eqnarray}
  P_x (k,r_\perp)&=& \left\langle \delta_{k,1} \delta^{*}_{k,2} \right\rangle,\\
  &=& \int_k^{\infty} \frac{q\,dq}{2\pi^2}
  P(q)\ J_0\left(r_\perp\sqrt{q^2-k^2}\right) \, .
\end{eqnarray}
where $\delta_k$ is the Fourier transform of matter over-density or
flux along a \emph{single skewer} and $r_\perp$ is the transverse separation
between skewers.

In Figure \ref{fig:ps1} we show the cross spectrum for the matter overdensity
and the corresponding linear predictions for both baryonic and baryon-less
universes for a few values of $r_\perp$.  Note that the acoustic feature not
only appears as a series of oscillations superimposed over a smooth shape,
but the oscillations have an increasing frequency.
Methods which involve measuring binned cross power spectra would need very
fine binning to avoid losing signal-to-noise.
While the full array of cross-spectra undoubtedly contains the acoustic
information, the difficulties of extracting it from this statistics appeared
daunting, and we do not consider the cross power spectra any further.

\subsection{Configuration space}
\label{sec:signal}

The configuration space statistics more naturally allow for complex
observing geometries, and irregularly sampled data while in principle
having more complex covariance properties.
We measured the correlation function of both $\delta$ and $\delta_f$ in
both real and redshift space by brute-force averaging over all possible
pairs of pixels in bins of perpendicular distance $r_\perp$ and parallel
distance $r_{||}$.

In real space, within the context of the peak-background split model one
can argue that the two-point flux statics should go as
\begin{equation}
  \xi_{ff}(r) = B(r) \xi_{\delta\delta} (r) + A(r),
\label{eq:qs}
\end{equation}
where $\xi_{ff}$ is the flux correlation function, $\xi_{\delta\delta}$
is that of the matter and $A(r)$ and $B(r)$ are functions that are
\emph{smooth\/} on large scales.
In practice we find that $B(r)$ is a constant and $A(r)$ is consistent with
zero in our simulations allowing us to write
\begin{equation}
  \xi_{ff}(r) = b^2 \xi_{\delta\delta}(r)
\end{equation}
with $b$ a large-scale bias.

In redshift space, we start by modeling the large-scale redshift space
distortions from super-cluster infall as \cite{1987MNRAS.227....1K}
\begin{equation}
  \xi(r, \mu) = \sum_{\ell=0,2,4} \xi_\ell(r)\,L_\ell(\mu),
\end{equation}
where $L_\ell(\mu)$ indicates the Legendre polynomial of order $\ell$ and
\cite{1992ApJ...385L...5H}
\begin{eqnarray}
  \xi_0(r) &=& C_0 \xi(r), \nonumber \\
  \xi_2(r) &=& C_2 \left[ \xi(r)  - \bar{\xi(r)}\right], \nonumber \\
  \xi_4(r) &=& C_4 \left[ \xi(r) + \frac{5}{2} \bar{\xi}(r) -
  \frac{7}{2} \widetilde{\xi}(r) \right] ,
\label{eqn:Kaiser}
\end{eqnarray}
with
\begin{eqnarray}
  C_0 &=& 1+ \frac{2}{3} \beta + \frac{1}{5} \beta^2 \nonumber, \\
  C_2 &=& \frac{4}{3}\beta + \frac{4}{7} \beta^2 \nonumber, \\
  C_4 &=& \frac{8}{35} \beta^2 \label{Cfacts}
\end{eqnarray}
and
\begin{eqnarray}
  \bar{\xi}(r) &=& \frac{3}{r^3} \int_0^r s^2 \xi_R(s) ds  \nonumber \\
  \widetilde{\xi}(r) &=& \frac{5}{r^2} \int_0^r s^4 \xi_R(s) ds .
\end{eqnarray}
We use $\xi_R$ here to explicitly denote the real-space correlation
function.  These expressions hold for both the matter and the flux, with
$\beta=f\equiv d\ln\delta/d\ln a\simeq \Omega_m^{0.6}\simeq 1$ in the
former case.  We leave $\beta$ as a free parameter for the flux,
although we expect it to be close to unity also.

\begin{figure}
  \includegraphics[width=0.9\linewidth]{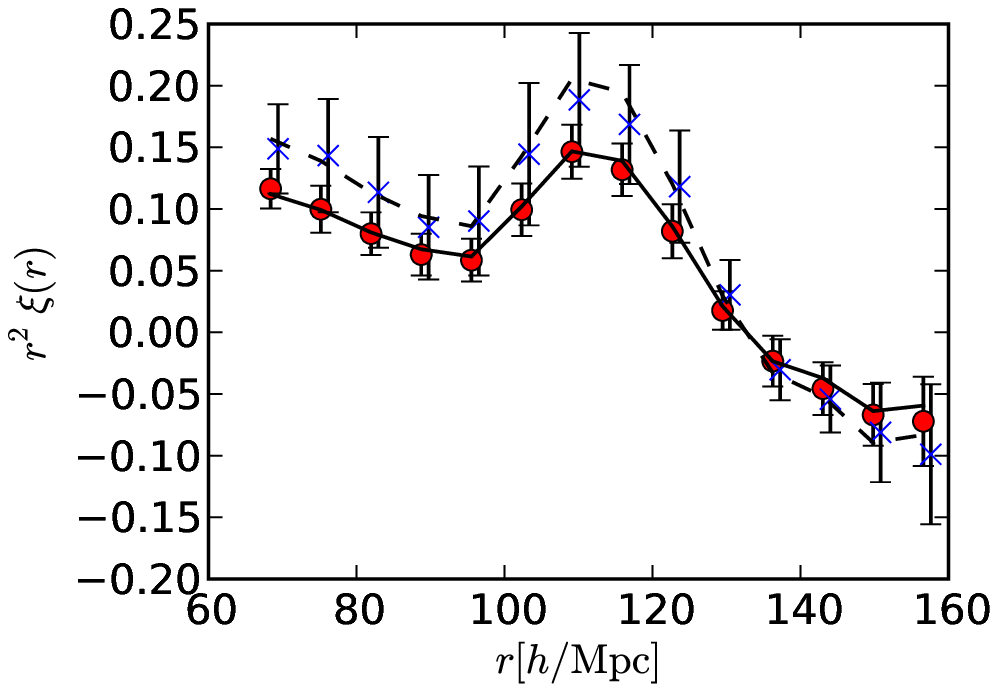} 
  \includegraphics[width=0.9\linewidth]{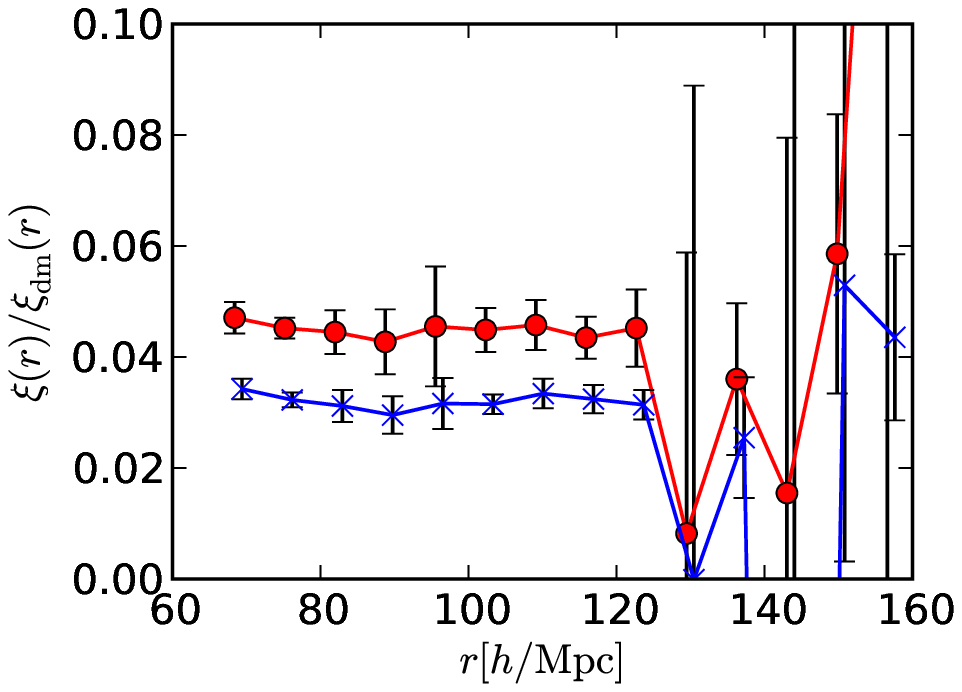} 
  \caption{(Top) The flux correlation function in real (filled red
    points) and redshift (blue crosses) space.  The black solid and and dashed
    lines are the scaled matter correlation functions. (Bottom) as
    above but divided by the equivalent matter linear correlation
    function.  There is no deviation from scale-independent bias seen
    in the lower panel.  Errors are estimated from the box-to-box
    scatter, are heavily correlated and only indicative of the
    underlying uncertainties.}
\label{fig:xiff}
\end{figure}

We show our main results in Figure \ref{fig:rxir}, which plots the
correlation functions (multiplied by $r^2$) in both real space and redshift
space, along with theoretical predictions from the simple model described
above.  The left panels show the real-space correlation function, where
one can clearly see the expected BAO `ridge' at $r\simeq 105\,h^{-1}$Mpc.
The middle panels show the correlation function in redshift-space, which is
very similar to the left panels with the exception of a bias.
The \lya flux follows the dark matter and shares the same $\beta\simeq 1$.
In the upper right-hand panel we show the predictions calculated using the
\emph{linear\/} theory correlation function and super-cluster infall with
$\beta=0.96$, appropriate to the matter at $z=2.5$.

Our ability to use linear theory to describe the acoustic signature at these
redshifts is an important feature of the method.
The dominant effect of non-linear clustering is to broaden the acoustic
peak, with an amplitude that can be estimated from the rms Zel'dovich
displacement \cite{ESW07,CroSco08,Mat08}.
At $z=2.5$ this is about $3\,h^{-1}$Mpc in our cosmology, to be compared to
the much larger intrinsic width of the acoustic feature (set by the diffusion,
or Silk, damping scale: $12\,h^{-1}$Mpc).
Adding these in quadrature we see non-linear evolution will only change
the peak width by $4\%$.
Thus linear theory should accurately describe the acoustic feature in the
matter at these redshifts.
It is also interesting to note that super-cluster infall does not generate an
elliptical contour in redshift space, as can be seen clearly in
Eq.~(\ref{eqn:Kaiser}).

Finally, the bottom right-hand panel of Figure \ref{fig:rxir} shows the
residuals $(\xi_{ff}-b^2\xi_{\delta\delta})/\xi_{ff}$ in redshift space.
The bright ridge is associated with the correlation function going though
zero, which increases the fractional errors, but there is no large-scale
radial structure in the difference plot.

We now focus our attention on the angularly averaged (or monopole) correlation
function, $\xi_0$, which in the large-scale limit should be proportional to the
real-space correlation function (Eq.~\ref{eqn:Kaiser}).
Figure \ref{fig:xidd} shows $\xi_0$ of the matter in real and redshift space,
using only information from the skewers, compared to the linear theory
predictions.  We note that linear theory describes the matter very well on
these scales and at this high redshift in both real and redshift space.

\begin{figure}
  \includegraphics[width=0.9\linewidth]{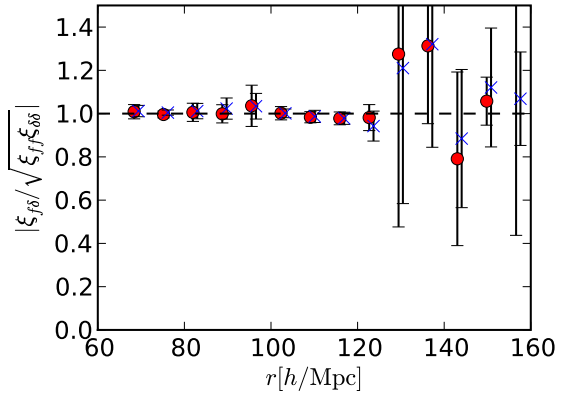} 
  \caption{The cross-correlation coefficient between flux and matter
     over-density in real (filled red points) and redshift (blue crosses) space.}
\label{fig:xicross}
\end{figure}

Figure \ref{fig:xiff} shows the real and redshift space correlation functions
of the flux and the ratio of the flux to matter correlation functions (the
`bias' squared).  The implied  bias is consistent with scale-invariant at
$b\simeq 0.2$, comparable to that obtained by \cite{2003ApJ...585...34M}.
Finally, Figure \ref{fig:xicross} shows the cross-correlation coefficient
between the flux and matter over-density,
\begin{equation}
  r\equiv \sqrt{\frac{\xi^2_{f\delta}}{\xi_{ff}\xi_{\delta\delta}}}
\end{equation}
which is unity when flux traces matter perfectly.  While in principle $r\le 1$,
when it is measured from simulations it can exceed unity due to noise.
The simulations show that the variations in the flux are tracing those in the
matter remarkably well on all scales of interest, in both real and redshift
space.  This validates the idea of measuring the BAO feature in the \lya
forest, and shows that there is no information lost in measuring flux rather
than matter fluctuations on these large scales even though the dynamic range
in the flux can be drastically smaller than in the density field.

\section{Fluctuating photo-ionization rate}
\label{sec:UV}

While our ability to measure the acoustic scale with galaxies has been
impressively demonstrated, we still do not know whether this will be
possible from the \lya forest.  The calculations above suggest that the
signal is present, and the propagation of statistical errors suggest the
measurement will be interesting \cite{2007PhRvD..76f3009M}, however there
are many systematic error which need to be controlled in order to obtain
the forecast statistical precision.

One possible cause for concern, in addition to systematic errors in the
measurements themselves, is non-gravitational contributions to the flux
correlations.  These could arise from hydrodynamic forces, radiative
transfer effects, reionization heating or other departures from the simple
FGPA assumed thus far \cite{2007arXiv0711.3358M}.
Many of these effects are expected to contribute mostly on small scales,
with no power preferentially on the acoustic scale, and by the arguments
provided earlier should not bias the BAO measurement
(see e.g.~Figure \ref{fig:xirescc}).
Diagnostics of these non-gravitational contributions can be found in the
higher moments of the flux \cite{2001ApJ...551...48Z,2004ApJ...606L...9F}
and all indications are that the forest is dominated by gravitational
instability on large scales.  However, the errors on such measurements are
still large enough that the issue is not settled.

One possible contributor to the observed power on large scales is fluctuations
in the UV background field or the photo-ionization rate ($\Gamma$).  Since
the attenuation length of the IGM at $z\sim 2-3$ is large, and the background
is thought to be dominated by rare sources (QSOs), $\Gamma$ may have spatial
structure on large scales.
If the IGM is in photo-ionization equilibrium the optical depth,
$\tau\propto\Gamma^{-1}$.
Since this is the most obvious source of `extra' large-scale power we
investigate this scenario in more detail.

To begin we estimate the expected magnitude of fluctuations in the
ionizing background field assuming it is generated by UV light from quasars
\cite{2004MNRAS.350.1107M,2004ApJ...610..642C,2005MNRAS.360.1471M}.
We do this numerically by populating each $(1.5\,h^{-1}{\rm Gpc})^3$ box with
mock quasars with luminosities following a broken power-law luminosity function
\begin{equation}
\Phi \propto \frac{1}{(L/L_\star)^\alpha+(L/L_\star)^\beta}
\end{equation}
with the parameters detailed in Table 5 of \cite{2004MNRAS.349.1397C}. 
We neglect any luminosity dependent conversion from optical of UV luminosity
in this preliminary study.
Each QSO is assumed to emit isotropically with constant luminosity $L$,
so the contribution to the photo-ionization rate from the $i$th QSO at
distance $r_i$ can taken to be
\begin{equation}
  \Gamma_i \propto L_i\ \frac{e^{-r_i/r_0}}{4\pi r_i^2}
\end{equation}
which neglects finite lifetimes or light-cone effects.  Here $r_0$ is
the 'attenuation length' of the IGM, which is $\mathcal{O}(100\,{\rm
  Mpc})$.  We chose $r_0=100\,h^{-1}$Mpc, close to the acoustic scale,
to maximize the potential contamination effect \cite{2004MNRAS.350.1107M}.

Since $r_0$ is large, we neglect any QSO clustering in our model, placing
the sources at random within the volume.
The spatial structure in the photo-ionization rate, or summed UV background,
depends on the luminosity function, in particular on the slopes at both the
faint and bright end.
Table \ref{tab:uvbg} shows some characteristic examples, with the variations
in the photo-ionization rate ranging from $31\%$ to $39\%$.
In what follows we shall take $0.01\le L/L_\star\le 100$, $\alpha=-3.31$ and
$\beta=-1.09$ as our fiducial model for the QSO component.
The correlation function of $\Gamma$ is close to constant at small scales,
and falls dramatically beyond the attenuation length, $r_0$.

\begin{table}
\begin{tabular}{ccccc}
$\alpha$&   $\beta$ &  $M_{\rm lo}$  & $M_{\rm hi}$ &    $\sigma_\Gamma/\langle\Gamma\rangle$ \\
\hline
-3.31 & -1.09  &   -29  &   -16  &   0.37 \\
-3.31 & -1.29  &   -29  &  -16  &   0.31 \\
\hline
-3.31 & -1.09  &   -29  & -20   &  0.38 \\
-3.31 &-1.29   &   -29  & -20   &   0.32 \\
\hline
-3.31 &-1.09   &   -29  & -22  &   0.39 \\
-3.31 &-1.29   &   -29  & -22  &   0.35
\end{tabular}
  \caption{Relative rms fluctuations in the UV background or
    photo-ionization rate, $\sigma_\Gamma/\langle \Gamma\rangle$, for
    different choices of the slopes in the QSO luminosity function and the
    magnitude limits of the QSOs we include.}
  \label{tab:uvbg}
\end{table}

To the particular realization of the QSO component we add a uniform piece (to
model the emission from faint AGN, galaxies and the IGM itself
\cite{2007arXiv0711.3358M}) such that the rms fluctuation of the total is
$5\%$, $10\%$ or $20\%$ and divide our original $\tau$ in every pixel by
the ionization rate at that location.
The overall normalization is rescaled to set the mean flux, $\bar{F}=0.8$.

\begin{figure*}
  \includegraphics[width=0.9\linewidth]{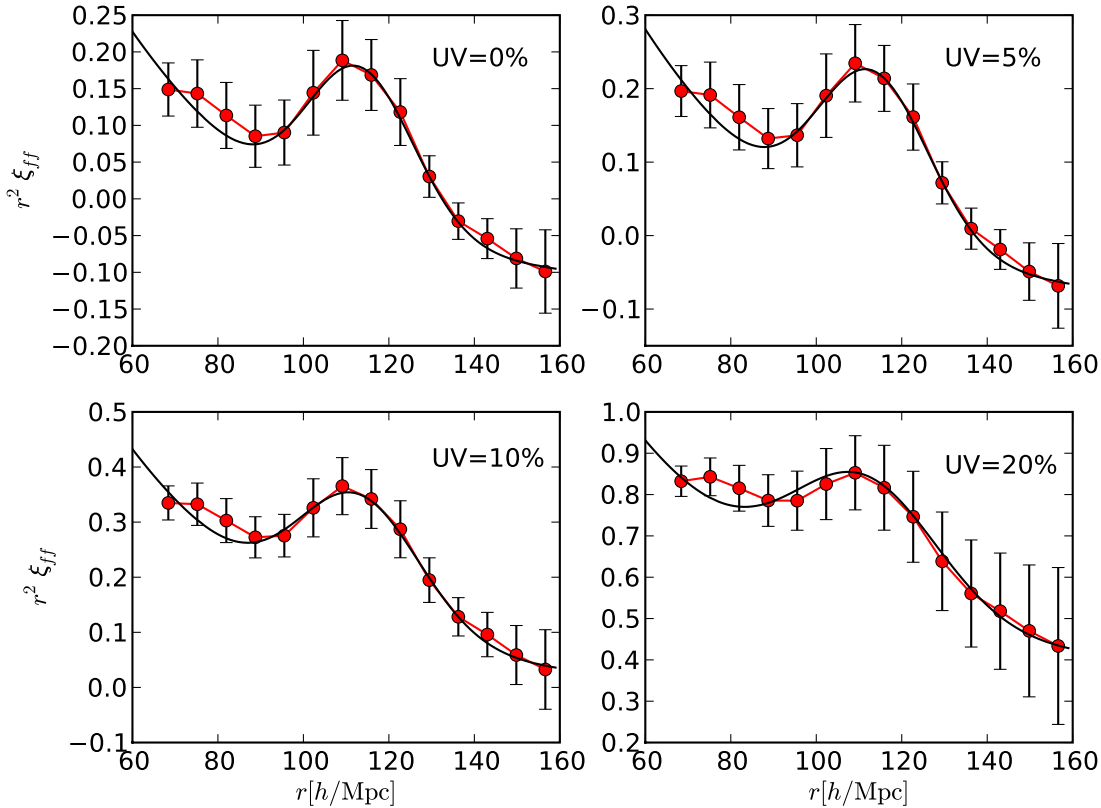} 
  \caption{The effect of photo-ionization rate fluctuations on the flux
    correlation function.  The four panels show correlation functions for
    UV fluctuations of 0\% (top left), 5\% (top right), 10\% (bottom left)
    and 20\% (bottom right).  Note that the vertical scale is different in
    each panel.  The solid lines correspond to the model of
    Eq.~\protect\ref{eqn:gaus}.}
    \label{fig:uv}
\end{figure*}

Figure \ref{fig:uv} shows the correlation functions of the resulting fluxes.
Fluctuations in $\Gamma$ significantly affect the correlation function on
large scales, increasing the level of power by a large factor.  The additional
power is however quite smooth, indicating that the acoustic information is
still accessible.  To see whether one can still reliably measure the position
of the peak in such degraded data, we have attempted to fit the data using the
following theoretical model
\begin{equation}
  \xi(r) = b^2\left(\xi_{\rm nb}(r) + \frac{h}{r^2} G(r_{\rm peak}, \sigma_{\rm
      peak})\right) + \mathcal{o},
  \label{eqn:gaus}
\end{equation}
where $b$, $h$, $r_{\rm peak}$, $\sigma_{\rm peak}$ and $\mathcal{o}$ are free
parameters, $\xi_{\rm nb}$ is the linear correlation function for a
\emph{baryon-less} universe, and $G(\mu,\sigma)$ is a Gaussian.
We do not advocate this as a realistic model of the correlation function,
but merely as a convenient prescription allowing us to asses whether there
are any biases introduced in the measurement of the position of the peak by
the presence of UV fluctuations.
We found best-fit point in each of the 8 boxes individually, assuming a diagonal
covariance with the plotted error bars, and used the scatter between these
best fits to asses the uncertainty in the parameters.  The results of these
fits are shown in Table \ref{tab:fit} and Figure \ref{fig:uv}. 

\begin{table}
\begin{tabular}{ccc}
model&   $r_{\rm peak}$ & $\sigma_{\rm peak}$ \\
\hline
dark matter, real space & $113.2\pm1.0$ & $16.5\pm2.5$ \\
flux, redshift space & $113.1\pm0.6$ & $17.9\pm2.9$ \\
flux, redshfit sp., UV=5\% & $113.1\pm0.6$ & $18.6\pm3.1$ \\
flux, redshift sp., UV=10\% & $112.8\pm0.7$ & $20.4\pm3.0$ \\
flux, redshift sp., UV=20\% & $111.1\pm1.1$ & $24.4\pm4.1$\\
\end{tabular}
  \caption{Fits of the correlation function, including UV background
    fluctuations, to the model of Eq.~\protect\ref{eqn:gaus}.}
  \label{tab:fit}
\end{table}

These results indicate that the peak position is not affected by the
presence of small UV fluctuations, but fluctuations larger than around
$\sim 20\%$ start to overwhelm the acoustic signal and biases begin to
introduced into the peak position recovery.  It is possible that more
sophisticated modeling could still recover the peak, but it is likely
that such large UV fluctuations will degrade the accuracy of the BAO
measurement.  In principle one can search for evidence of UV
fluctuations in existing QSO spectra.  We also find a weak evidence
that the width of the peak broadens as UV fluctuations are increased.

\section{Discussion and Conclusions}
\label{sec:conclude}

Baryon acoustic oscillations have become one of our most promising
methods for determining cosmological distances and hence the expansion
history of the Universe.  The structure in the spectrum of distant
quasars, which is thought to trace the structure of the IGM at near
mean density, has been suggested as a relatively cheap method for
measuring BAO at high redshift \cite{2003dmci.confE..18W}.  It is
difficult to compute the BAO signal in the \lya forest analytically,
since it involves projection of a non-linear mapping of a non-linear
density field in redshift space.  Instead we have used large-volume
N-body simulations.  We argued that while the small-scale physics in
these simulations is not accurate, the overall picture should not
change drastically with more realistic simulations.  Our conclusions
can be summarized as follows.

We see clear evidence for the acoustic scale in our 180,000 mock spectra.
While the BAO signal is present in both Fourier and configuration space
statistics, the latter seem to be the better for analyzing the data.
There are two mains reasons for this.  First, the BAO feature is a single
comparably narrow feature in the correlation function, rather than a series
of oscillations in the power spectrum.  Second, the complex nature by which
the \lya forest samples the underlying field makes the analysis considerably
more subtle in Fourier space, or conversely the mask much easier to handle
in configuration space.

Within the peak-background split approximation, the \lya flux follows the
mass fluctuations on large scales.  Within the FGPA the flux is a highly
non-trivial but deterministic transformation of the underlying density field
and we find that the \lya flux correlation functions trace the mass with
high fidelity on large scales, with negligible scale dependent bias.
We also find that the cross-correlation between the flux and matter is
consistent with unity.   This is both surprising, given that the flux has
limited ``dynamic range'' compared to the mass, and very encouraging. 

Given the difficulty of observing in the deep ultra-violet bands from
the ground, \lya observations are limited to redshifts $z>2$ from
ground based observations. The upper practical limit is set by the
fact that quasars must be bright enough to observe them with a survey
instrument and this limits the upper redshift to $z\sim2-3$. Our
simulation redshift of $z=2.5$ is thus at a typical value for a future
instrument such as BOSS. Quantitative details are likely to change
with redshift, but our main qualitative conclusions are robust,
because there is no fundamental change in the properties of the
intergalactic medium in the redshift range of interest.

Similarly, our skewers were simulated at a fixed redshift, while the
real \lya forest probes an evolving intergalactic medium. This will
manifest itself as an effective large-scale bias that evolves with
redshift, but is unlikely to change abruptly. We therefore propose to
analyse the data in discrete redshift bins that are wide enough to
have enough Fourier modes along the line of sight, but are at the same
time narrow enough to allow an approximation of a constant or linearly
varying effective large-scale bias.

However, a number of other systematic effects will need careful
investigation before the BAO program can be carried out with \lya
forest data.  We have begun by investigating the effect of spatial
structure in the photo-ionization field, which modulates the optical
depth on scales comparable to the acoustic feature.  We find that such
a modulation does not introduce any confusion into the determination
of the acoustic scale, though it does modulate the total amplitude of
the flux correlation function.  This is not too surprising, the
photo-ionization rate does not know to add power at $105\,h^{-1}$Mpc
differently than at $104$ or $106$, but is also encouraging.

As the simulations become increasingly realistic we may begin to see some
scale-dependence in the bias.  We therefore suggest that one should model
BAO in the \lya forest, by measuring a set of bands in the correlation
function, together with a linear $\beta$ parameter.
This could for example be achieved using an (optimal) quadratic estimator.
The resulting $\beta$ should be close to that of the matter, i.e.~close to
unity.
In order to be conservative, the resulting correlation function could be
compared with the theoretical predictions for the matter correlation
function using the formalism in Equation \ref{eq:qs} with $A(r)$ and $B(r)$
being smooth, one or two parameter functions.
Sufficiently accommodating functions can also protect against unexpected
systematics.

Significantly more work needs to be done before this program can be executed
and there are several important effects that need to be understood.
First, quasar continuum fluctuations can potentially give rise to large-scale
fluctuations that might have a preferred scale.  The fact that we are
cross-correlating different spectra alleviates the problem to some extent,
but may not solve it.
Similarly, metal contamination can also give rise to spurious correlations.
If fluctuations in the UV background, or photo-ionization rate, are large
they may imprint large-scale power in the flux correlation function which
can overwhelm the acoustic signal and reduce the sensitivity of the
measurement.

Many of these issues can be dealt with using two basic approaches.
The first is to model them.  For example, pixel pairs which are thought
to be contaminated by a metal doublet could be `blinded' within an
optimal quadratic estimator.
A second method is to split the full sample into all possible subsamples.
For example, quasars of the same absolute magnitude or the same source
redshift are more likely to have similar continuum fluctuations and one
could check the results by splitting the sample into subsamples according
to QSO magnitude or redshift.  
Our results provide added impetus to further develop these promising
ideas.

\vspace*{1cm}

\begin{acknowledgments}
  Authors acknowledge useful discussions with members of the BOSS \lya
  working group and Uro\v{s} Seljak. AS is supported by the Berkeley
  Center for Cosmological Physics.  MW is supported by the NSF and
  DoE.  The simulations presented in this paper were carried out using
  computing resources of the National Energy Research Scientific
  Computing Center and the Laboratory Research Computing project at
  Lawrence Berkeley National Laboratory.

\end{acknowledgments}

\bibliographystyle{apsrev}
\bibliography{lyaone,cosmo,cosmo_preprints}

\end{document}